\documentclass[a4paper]{jpconf}
\usepackage{graphicx}
\usepackage{epstopdf}

\def\be{\begin{equation}}
\def\ee{\end{equation}}
\def\bea{\begin{eqnarray}}
\def\eea{\end{eqnarray}}

\begin{document}
\title{Black hole gas in TeV-gravity models}

\author{M\'onica Borunda and Manuel Masip}

\address{CAFPE and Departamento de F\'isica Te\'orica y del Cosmos, 
Universidad de Granada, E-18071 Granada, Spain}

\ead{mborunda@ugr.es, masip@ugr.es}

\begin{abstract}
In a plasma at 
temperature close to the fundamental scale a small fraction  
of particles will experience transplanckian collisions that may 
result in microscopic black holes (BHs).  We study the dynamics 
of a system (a {\it black hole gas})
defined by radiation at a given temperature coupled to a 
distribution of BHs of different 
mass. Our analysis includes the production of BHs in photon-photon 
collisions, BH evaporation, the absorption of radiation, 
collisions of two BHs to give a larger one, and the effects of the
expansion. We find that the system may follow two different
generic paths depending on the initial temperature of the plasma.
\end{abstract}

\section{Introduction}
The presence of extra dimensions may accommodate
a fundamental scale of gravity 
$M_D$ much lower than the Planck mass 
$M_{P}\sim 10^{19}$ GeV
\cite{ArkaniHamed:1998rs,Randall:1999ee},
with $M_D\approx 1$ TeV solving the so-called
hierarchy problem. 
If that were  the case, collisions of two particles 
at center of mass energies above
$M_D$ would be dominated by 
gravity. In particular, one expects that at small impact 
parameters gravity {\it bounds} the system and the particles
collapse into a microscopic BH. 
This process, extensively studied as a possibility for 
the LHC, would be also relevant in the early universe if
the initial temperature $T$ was ever close to the TeV scale.

At such temperatures two particles in the high-energy tail 
of the distribution may experience a transplanckian 
($\sqrt{s}>M_D$) collision and produce a mini-BH. If the 
BH is colder than the environment ($T_{BH}<T$) 
it will gain mass, and as it grows 
its temperature drops further. This would distinguish
BHs from other massive particles or string excitations that
may also be produced at high $T$: the heavier the BH, the 
colder and more
stable it becomes, whereas heavier elementary particles 
have a shorter lifetime. BHs seem indeed a key ingredient
in the dynamics of a plasma at temperatures close
to the fundamental scale. Notice that the system
is peculiar in the sense that 
the heat flows from the hot plasma to the colder BHs,
but the effect is to cool the BHs.

In the usual TeV-gravity models with flat extra dimensions 
(ADD) a high $T$ seems unconsistent 
with the standard cosmology \cite{Hannestad:2001nq}.
There the KK excitations of the graviton have
an effective 4-dim mass proportional to
$m_c\equiv L^{-1}$ and very weak couplings
($\approx \sqrt{s}/M_P$) 
to ordinary matter. Even if the initial configuration 
consists of particles only in the 
4-dim brane, at $T\gg m_c$ these bulk gravitons 
will be abundantly produced (due to
their large multiplicity) in annihilation
of brane particles. If $T$
is large these (long-lived) 
gravitons will change the expansion
rate at the time of primordial nucleosynthesis. Even if 
$T$ is initially as low as 1 MeV, their late decay
will distort the cosmic background radiation in an unacceptable
way. Obviously, a temperature close to $M_D$ 
would bring too many massive gravitons.
In contrast, warped (RS) models predict KK gravitons 
just below $M_D$ with unsuppressed couplings, so
a few resonances are enough to give an order one gravitational coupling
at $M_D$. In RS models the 
bulk is basically empty at $T<m_c\approx 0.1 M_D$.

Here we will consider
{\it hybrid} models \cite{Giudice:2004mg}
with $n$ extra dimensions and  
two uncorrelated parameters: the effective 
compactification scale $m_c$
and the fundamental scale $M_D$. We will assume a mechanism
that {\it pushes} the KK gravitons towards the 4-dim brane and
increase their coupling to matter: 
\be
{s\over M_P^2}\rightarrow {s\over M_D^2}
\left( {m_c\over M_D} \right)^{n}\;.
\ee
In this way a smaller number of KK modes will imply an
order one gravitational interaction at the same scale 
$\sqrt{s}=M_D$. 
If the {\it free} parameter $m_c$ takes the value 
$m_c\approx  M_D ( M_D/ M_P )^{2/n}$ 
we recover ADD, whereas for $m_c$ approaching
$M_D$ we obtain RS. At distances below 
$1/m_c$ gravity would be ($4+n$)-dim (similar to ADD)
whereas at larger distances it becomes 4-dim (like
in the usual RS).
The KK gravitons are not 
produced at $T<m_c$, while their larger couplings to matter
decouple them from the plasma at $T$ below their mass 
and make their lifetime shorter. This framework with
$m_c>1$ GeV avoids all 
astrophysical \cite{Hannestad:2003yd} and
cosmological \cite{Hannestad:2001nq} bounds for 
any value of $n$.

There are two basic points that distinguish the TeV-gravity 
scenario from the standard 4-dim one. First, BHs 
are colder and live longer here than in 4-dim. 
Second, in these models the Hubble time 
$H^{-1}\approx 1/\sqrt{G_N \rho}$ 
at the temperature where
the BHs are produced is much longer (in terms of the 
fundamental time scale) than in four dimensions:
\be
H^{-1}_{(4)}\sim \frac{1}{M_{Pl}}\, , \,\,\,\,\,\,\,\,\,\,\,\,\,
H^{-1}_{(4+n)}\sim\frac{M_{Pl}}{M_D}
\left( \frac{m_c}{M_D} \right)^{n/2}
\frac{1}{M_D}
\ee
where we have assumed a bulk in thermal equilibrium with
the 4-dim brane and $\rho\approx T^{4+n}/m_c^n$.
As a consequence, in a $(4+n)$-dim universe there is plenty 
of time for BHs to be produced and grow, changing the dynamics 
of the early universe. In a 4-dim universe 
the temperature drops much faster, the BHs
of mass near the threshold $M_P$ get then hotter 
than the environment and evaporate in a time $\approx 1/M_P$.

\section{One black hole}
Let us first consider a single BH in an expanding universe. 
We take $M_D= 1$ TeV, and $m_c = 10$ GeV, with initially the 
brane and the bulk in thermal equilibrium at 
a temperature $T=T_0$.
Our hybrid model at energies above $M_D$ and distances 
smaller than $1/m_c$ is strongly coupled and ($4+n$)-dim, 
so the BH is initially a regular
($4+n$)-dim one. If it is colder than the plasma ($T_{BH}<T_0$)
it grows, and when it reaches 
a radius $r_H=1/m_c$  it fills up the whole compact space.
At this point the BH keeps gaining mass,  but its 4-dim size 
does not increase because gravity is strong only up to distances of
order $1/m_c$ (the inverse mass of the lightest KK graviton).
Beyond this distance gravity is 4-dim, since 
only the massless graviton
is effective. Therefore, the BH grows at constant radius and
temperature $T_{BH}\approx m_c$ until its mass
reaches $M\sim M_P^2/m_c$, when it becomes purely 4-dim. 
At $T>m_c$ the BH mass changes according to
\be
\frac{dM}{dt}=\sigma_4A_4(T^4-T_{BH}^4 )+
\sigma_{4+n}A_{4+n}(T^{4+n}-T_{BH}^{4+n})
\label{dMdt}
\ee
where the negative terms account for the evaporation into
brane and bulk species and in $A_{4+n}$ (the BH area) we
take a maximum radius of $1/m_c$ along
the extra dimensions.

The expansion rate 
(we assume that the extra dimensions are 
frozen) of the universe is dictated by the
radiation energy density in the bulk and the brane, 
\be
\rho_{rad}\sim \frac{\pi^2}{30}T^4\left( g_\star +
g_bc_n \frac{T^n}{m_c^n}\right)\, .
\ee
At $T>m_c$ the bulk energy dominates 
and the temperature decreases as $T\sim t^{-2/(4+n)}$ 
with time, whereas 
at $T<m_c$ all the bulk energy is 
transferred to the brane and the universe expands 
like in the standard 4-dim cosmology.

In figure 1 we show the evolution of a BH  of mass 
$M=100$ TeV with $M_D=1$ TeV, $n=1$, just photons 
($g_\star =2$) and gravitons ($g_b = 5$) in the bulk 
at $T_0 = 100,200$ GeV.

\begin{figure}[h]
\begin{minipage}{13.6pc}
\includegraphics[width=14pc]{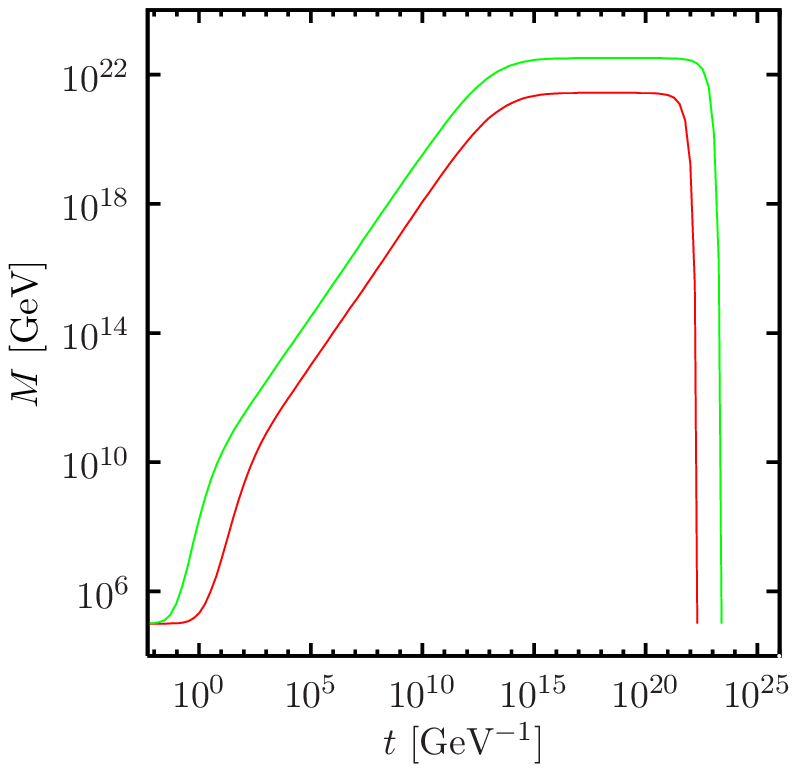}
\caption{\label{label}Mass of a single BH.}
\end{minipage}\hspace{2pc}%
\begin{minipage}{14.2pc}
\includegraphics[width=14pc]{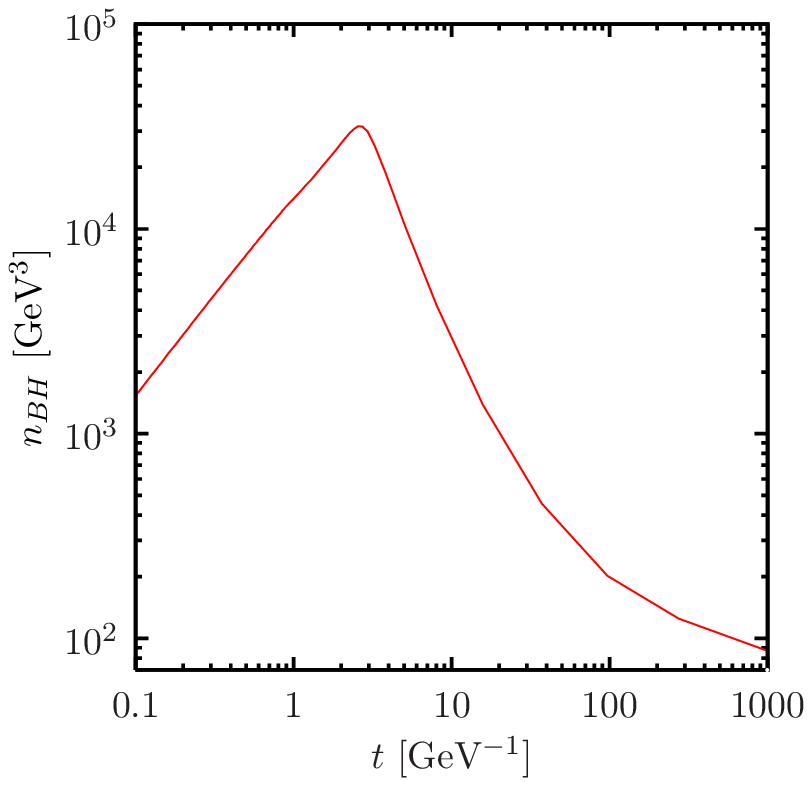}
\caption{\label{label}$n_{BH}$ at $T_0=200$ GeV.}
\end{minipage} 

\end{figure}

\section{Black hole gas} 

Let us now simulate the evolution of the system, a {\it black hole gas},
including the different processes
that may take place:
 \begin{itemize}
 \item  Black hole formation in collisions of two particles.
 \item Collisions of two BHs into a larger one. 
  \item Absorption and emission of radiation.
 \item  Effect of the expansion of the universe.
  \end{itemize}

We will consider an initial configuration with only radiation
at $T_0$ both in the brane and the $n$ dimensional bulk.
Particles with enough energy will collide to form BHs that 
will grow if $T_{BH}<T_0$. 
If $T_0$ is smaller than $M_D$ then the BHs 
are always non-relativistic matter, with a 
kinetic energy ($\approx T$) much smaller than the
mass. We take $E_{BH}\sim M_{BH}$ and 
$v_{BH}=\sqrt{2 T / M_{BH}}$.

We will assume that the BH gas is a two-component 
thermodynamic system formed by radiation with 
a time dependent temperature $T(t)$ and BHs described 
by a distribution $f(M,t)$. 
$f(M,t)$ describes the number of BHs 
of mass $M$ at a given time $t$ 
per unit mass and volume. In this way, 
the energy density of the universe is the addition of 
$\rho_{rad}$ (for brane and bulk radiation) 
plus $\rho_{BH}$ (the matter density in the BHs).

As the radiation cools down due to  the absorption and the expansion, 
lighter BHs (of 
higher temperature) become hotter than the plasma and evaporate.
This reheats the universe and provides energy to be absorbed
by the larger BHs. We are interested in the evolution of
the system. Does $\rho_{BH}$ ever dominate the energy density 
of the universe? How big the BHs grow? 
How much the final configuration depends on the 
initial temperature of the universe?

\section{Results}
 
We have identified 2 generic scenarios depending 
on $T_0$. Larger values of $T_0$ (above $\approx 0.2 M_D$)
produce {\it many} BHs, which grow fast and dominate
$\rho$. In this case the BH gas goes through four phases 
\cite{Borunda:2009wd}:
 \begin{itemize}
 \item BHs of mass above a critical mass 
corresponding to $T_{BH} = T_0$ are produced (see Fig.~2) and absorb 
radiation. As their number grows collisions between 
two BHs become important.
BHs dominate the energy density very fast 
($t\approx 10^{-12}H^{-1}$),
reducing the temperature of the radiation.
 \item As $T$ drops BH production  
stops. Light BHs become hotter than the plasma and 
evaporate. This reduces the number of BHs but the 
average BH mass keeps growing, since the colder (bigger) BHs 
keep absorbing radiation. Once a BH reaches a mass 
around $10^7$ GeV its radius $r\approx 1/m_c$ stops growing.
 \item When $T$ drops to $T_{BH}\approx m_c$, the slow 
absorption of radiation by the heavier BHs is 
compensated with the evaporation of the lighter ones,
keeping $T$ basically constant. The energy density is still 
dominated by BHs.
 \item At times of order $H^{-1}$ the 
expansion cools the radiation. The BHs decay fast and 
the universe becomes radiation dominated. Likewise, the 
lightest KK modes also decay fast and only 4-dim photons 
survive below $T\sim 0.1 m_c$.
 \end{itemize}

In the case with a lower initial temperature  
($T_0$ below $\approx 0.1 M_D$)
the basic difference is that the 
BHs do not dominate the energy density at any
time. In this second case we distinguish 
three phases:
\begin{itemize}
\item BHs are produced at a constant rate but since 
their number is so small 
collisions between them are negligible. These BHs 
grow like the single BH described in Fig.~1.
\item At a Hubble time the expansion cools the radiation
and stops BH production. The average 
BH mass keeps growing, but the universe is always 
radiation dominated.
\item Due to the expansion the temperature of the 
universe drops below the temperature of the BHs, which
 evaporate in a time scale that depends on
the ratio $M_D/m_c$ (it is around $t\sim 10^{22}$ GeV$^{-1}$
for $M_D/m_c=10^2$).
\end{itemize}
Both scenarios seem consistent with primordial 
nucleosynthesis. In the second case the BHs are a small 
contribution to $\rho$ that would not change the expansion rate of the 
universe. For larger values of $M_D/m_c$ one may obtain 
BHs whose late decay may introduce distortions in the 
diffuse gamma-ray background, or even BHs with a lifetime 
larger than the present age of the universe. These ones
could account for additional dark matter or become 
seeds for macroscopic primordial BHs.

We think that the work presented here is a first step
in the search for possible observable effects of these
BHs and on the understanding
of the initial conditions in the early universe.

\section*{References}

\smallskip

\end{document}